\documentstyle[11pt,paspconf,epsf]{article}
\begin{document}
\title{Evolutionary population synthesis: the effect of binary systems}

\author{J.M. Mas-Hesse, M. Cervi\~no}
\affil{LAEFF-INTA, P.O. Box 50727, E-28080 Madrid, Spain.}

\begin{abstract}
  
We present in this contribution our set of multiwavelength synthesis models
including the evolution of single and binary stars. 
The main results we have obtained can be summarized as follows: (a) massive
close-binary systems will start to experience mass transfer episodes after
the first 4 Myr of the starburst evolution; (b) as a result of these mass
transfer processes, stars of relatively low initial mass can loose
completely their envelope and become a WR. In this way, the formation of WR
stars is extended over longer than 15 Myr, and does not stop at 6 Myr as
predicted by models including only single stars; (c) WR stars can thus be
coeval with red supergiants, which peak at around 10 Myr for solar
metallicities; (d) the accretion of mass will originate relatively massive
stars at ages for which they should have already disappeared; these stars,
together with the  WR stars formed in rather evolved clusters,
increase the production of ionizing photons, so that the H$\beta$
equivalent width will 
not drop as rapidly as predicted by models considering only individual stars;
and (e) the mass transfer to compact companions will produce an
additional source of high-energy radiation in the form of high-mass X-ray
binaries, not predicted either by standard synthesis models.

\end{abstract}


\section{Introduction}

In the last years convincing observational evidences have been collected 
about the presence of starburst regions in or around active
Seyfert~2 nuclei (Heckman et al. 1997; Gonz\'alez-Delgado et al. 1998, 
and references therein). It has been found that most of the UV light 
from these objects originates in (circum-)nuclear star formation sites; 
the possible connection between the nuclear activity and the properties 
of these starbursts is nevertheless still a matter of debate. 
According to the unified schemes of Seyfert galaxies, the active nucleus
should be hidden by an opaque torus in the case of Seyferts~2, explaining
so why the UV light collected is indeed dominated by these young, massive
stars. The low degree of contamination by the active source in the UV range
allows therefore to get detailed information about the properties of the
star formation processes. Moreover, extrapolating to the radio -- X-ray
ranges the emission associated to the starbursts, it should be possible to
disentangle the fractional contribution of both sources (the starburst and
the active nucleus) at different energy ranges.  Multiwavelength evolutionary
synthesis models normalized to the observed UV emission would be the ideal
tool to perform this analysis.

With these ideas in mind we started some years ago a program to extend our
evolutionary synthesis models to high energy ranges (soft and hard X-rays,
gamma-rays). Preliminar results were presented in Cervi\~no et
al. (1996). It was clear that to properly reproduce the high energy
emission, the effects of mechanical energy release (by stellar winds and
supernova explosions) on heating the diffuse interstellar gas had to be
included, as well as the evolution of binary systems. 
 The process of mass transfer between close interacting
companions can affect the evolution of the individual stars, and can also
trigger high energy emission when one of the stars has already evolved to a
compact object. Binaries have also been included in starburst computations
by Vanbeveren et al. (1997) and Van Bever \& and Vanbeveren
(1998). Schaerer \& Vacca (1998) performed also a simple approximation to
the effects of binary systems on the evolution of young starbursts. In a
recent review (Vanbeveren et al. 1998) the detailed description of the
evolution of massive close binary systems has been presented. This has also
been summarized by Vanbeveren et al. in several contributions in this
volume.

We present in this contribution our set of stellar population models
including the evolution of single and binary stars. We will stress the
effects of binary evolution on the stellar population structure. 
In Sect.~2 we present our set
of evolutionary synthesis models and in Sect.~3 we discuss the predictions
on the stellar population.  In an accompanying contribution in this volume
(Cervi\~no \& Mas-Hesse, {\em ``Hard X-ray-to-radio energy distributions in
starburst galaxies''}), the multiwavelength spectral energy distribution
and the effects on the H$\beta$ equivalent width and HeII emission line
are discussed. The complete set of models will be published elsewhere. 

\section{Evolutionary synthesis models for single and binary stars}

The basic ideas about our evolutionary synthesis models for {\em single}
stars have been already discussed in Mas-Hesse \& Kunth (1991, hereafter
MHK) and in Cervi\~no \& Mas-Hesse (1994, hereafter CMH).  Basically, our
models are based on different Initial Mass Function (IMF) slopes 
($\alpha$= 1,
2.35 --Salpeter-- and 3), two extreme star formation regimes (Instantaneous
--IB-- and Extended Bursts --EB--) and five sets of stellar evolutionary tracks
with different metallicities taken from the Geneva group (Schaller et
al. 1992 and references therein). In this section we discuss the framework
we have used to include the evolution of binary systems, which is based on
the prescriptions of Vanbeveren (1991) and Vanbeveren et al. (1998),
somewhat modified and simplified.

We assume that the stars evolve like single, non-rotating stars following
their theoretical tracks until mass transfer episodes or supernova
explosions occur. The procedure for track interpolation in mass has been
discussed in detail in CMH. We want to remark that we compute the evolution
of each individual mass along the HR diagram, and obtain at each time step
the total number of stars and their luminosity for each spectral type and
luminosity class bin.
In case of the stars forming a binary system, we recompute continuously
their mass ratio and orbital separation using their instantaneous masses,
as defined by the stellar tracks. We have assumed circular orbits along
the evolution of the systems, although we have also considered the possible
formation of very eccentric systems after a supernova explosion. 

\subsection{Initial distributions}

The parameters determining the evolution of binary systems are mainly the
mass of each component, their mass ratio $q$ ($= M_2/M_1$ where $M_1$
corresponds to the more massive star) and the orbital semi-axis, $a$ or
period $P$, together with the fraction of stars formed in such systems. The
initial distributions of these parameters will define the evolution of the
young clusters and their observational properties.

\begin{table}
  \label{tab:input}
  \begin{tabular}{lll}
    \hline
    Input parameter         & Law           & Method \\
    \hline
    IMF & $dN/dm \propto m^{-\alpha}$ ($\alpha$= 1, 2.35, 3)&Montecarlo\\
    & Mass limits: 2-120 M$_\odot$ &\\
    Star Formation Regime & Instantaneous, extended & \\
    Binary frequency & 50\% (10, 30, 70, 90\%)&\\
    Mass ratio  & $q = M_2/M_1$ & Montecarlo \\
    Orbital separation & $dN/da \propto a^{-1}$  & Montecarlo \\
    & $\frac{a_{min}}{R_\odot} < 6(\frac{M_1}{M_\odot})^{1/2}$, 
      $\frac{a_{max}}{a_{min}} = 5000 $ & \\
    \hline
  \end{tabular}
  \caption[]{Initial input parameters distributions used in the models.}
\end{table}

We have assumed by default that around 50\% of the stars are formed in
binary systems, supported by the frequency of WR stars in binaries obtained
by Vanbeveren \& Conti (1980) for the solar neighbourhood. Nevertheless, we
have also explored the effect of assuming different binary frequencies (10,
30, 70 and 90\%). In general, even assuming low frequencies around 10\%,
the effects on the evolution of the stellar population are already
remarkable. We want to stress that this {\em binary frequency f} refers to
the total number of binary systems, and not just to those experiencing mass
transfer. Binary interactions will affect indeed to only less than 5\% of
all systems (Maeder \& Meynet 1994).

The Initial Mass Function is generated by a Montecarlo technique, as
explained in Arnault (1990), which provides an initial mass value for each
star formed in the cluster. It is forced to follow a power-law with 3 fixed
values of the slope. The mass ratio distribution has been obtained from the
same Montecarlo run, by randomly associating two subsequent mass values,
only constrained to comply with the defined binary frequency. Finally, the
orbital separation distribution has been also generated by a Montecarlo
routine, forced to a $dN/da \propto a^{-1}$ function. In order to have
statistical significance, we have used a distribution of 5$\times$10$^5$ stars
(i.e. 1.25$\times$10$^5$ binary systems). As explained in CMH, smaller
distributions of stars can lead to significant stochastic variations in the
stellar population from one Montecarlo run to another. While this feature
reproduces what we observe in small clusters, for the purposes of this work
we have assumed only large numbers of stars with well defined initial input
distributions.
The initial distributions we have assumed are summarized in 
Table~1.  

  \begin{figure}
    \plotfiddle{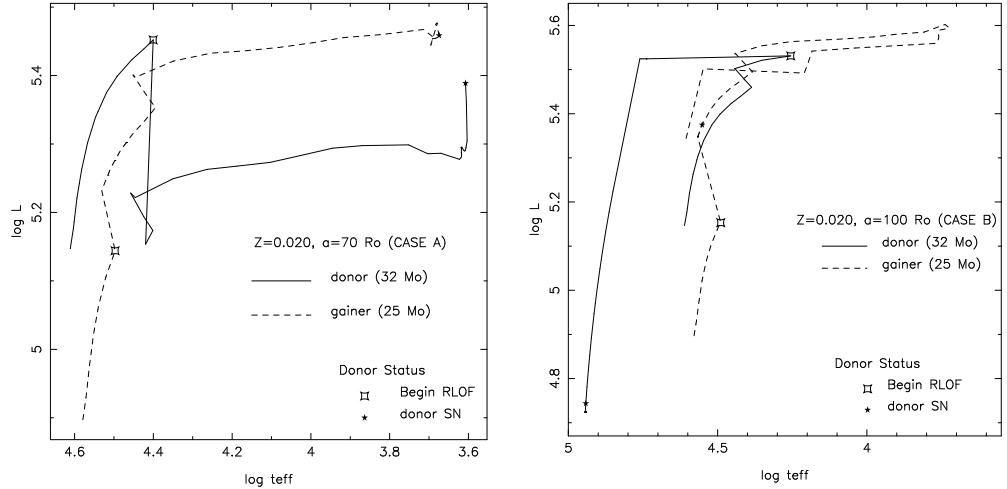}{6cm}{0.}{70.}{70.}{-190.}{0.}
    \caption[]{Evolution of a binary system (a) following Case A 
      and (b) following Case B of mass transfer.}
    \label{caseA}
  \end{figure}  

\subsection{Roche Lobe Over-Flow phase (RLOF)}

Mass transfer episodes during Roche lobe overflow are the most critical
process when analyzing the evolution of binary systems. We test
continuously whether RLOF episodes occur by comparing the Roche lobe size
with the stellar radii at each time step.  Vanbeveren et al. (1997) show
that systems in which the primary has an initial mass larger than 40
M$_\odot$ with periods such that RLOF should start during or after the LBV
phase, can avoid the mass transfer episode due to the loss of mass via
stellar winds, so that no interaction will occur. We have therefore
neglected possible mass transfer episodes in stars initially more massive
than 40 M$_\odot$ or when the primary has reached the WR phase when RLOF
should occur. This restriction has important consequences, in the sense
that the properties of a cluster containing binary systems will not differ
significantly from a single-stars-only cluster during the first Myrs of
evolution, until the most massive stars in the clusters explode as
supernovae and might originate binary systems with a compact companion. 
When RLOF occur, the evolution of the donor and gainer remnants will depend
mostly on their initial mass, their evolutive status, and the amount of
mass transferred:

\subsubsection{$\bullet$ Donor star.}
The donor star will generally loss its outer layers during mass transfer
episodes. Depending on its structure and evolutive status we can identify 3
main cases:

\renewcommand{\theenumi}{\roman{enumi}}
\renewcommand{\labelenumi}{\theenumi.}
\begin{enumerate}
\item {\it The Donor is an H burning star (Case A).} In this situation 
mass transfer will last until H is exhausted in the
nucleus of the donor star, so that the transfer episode will be extended
over the remaining main sequence lifetime of the star. As a result, the
donor lifetime will be longer that that of a single star with the same
initial mass. As a net effect of Case A episodes the mass ratio of the
system will be reversed, and the individual stars will continue their
evolution according to their new mass, considering that around 50\% of the
mass has been assumed to get lost during the transfer episode. We show in
Fig.~\ref{caseA}a the evolution of a system experiencing Case~A mass
transfer.

\item {\it Donor between H and He burning phase (Case B).}
In this case, the star looses mass until the atmospheric H abundance
is around 0.2 (de Greve \& de Loore 1992).  The remnant mass will be,
according to Vanbeveren et al. (1998):

  \begin{equation}
    m_d^{post~RLOF} = 0.093 \times (M_d^{pre~RLOF})^{1.44}
  \end{equation}
  
As a first approximation we have assumed this law for all metallicities
considered, but it may change with metallicity as discussed by Vanbeveren
et al. (1998). If the remnant mass of the donor star is larger than 
5~M$_\odot$, it becomes a WR-like star, according to Vanbeveren et
al. (1997). Bare He burning stars formed after RLOF and with less than 
5~M$_\odot$ have been
counted also as Wolf-Rayet stars, although their expected weaker, but
non-zero, stellar winds might not be able to form a thick mantle originating
the spectral features associated with WRs. The value of this mass limit
will be refined in the future, when more precise atmosphere models for this
kind of stars become available. In the meantime we want to stress that the
WR population our models are predicting at evolved ages (longer that
11~Myr) does not necessarily show the same spectral features than more
massive WRs. 

In order to follow the evolution of the star after mass transfer, we have
assumed that: the evolutive status remains the same (no time correction); 
the surface abundances are equal to the core ones;
the instantaneous mass is recalculated at each time step taking into
    account the remnant mass and the mass loss laws for the different WR
    phases; 
the WNL phase will last until all the H is removed from the
    envelope, assuming a mass loss rate of $4 \times 10^{-5}$ M$_\odot$
    yr$^{-1}$ and        
the WC phase will follow the mass-luminosity and mass-radius relations
    from Schaerer \& Maeder (1992) for WR stars and Langer (1989) for
    He stars (mass after RLOF lower than 5 M$_\odot$).
In Fig~\ref{caseA}b we show the evolution of a system experiencing a Case~B
RLOF episode.

\item {\it Donor in He burning or later phases (Case C).}
In this case the effects of shell-burning regions in the structure of the
star is not clear, so that the evolution after RLOF is quite
uncertain. Anyway, since the donor remnant will be already very evolved,
its remaining lifetime will be very short and the number of these systems
will be a minority (Vanbeveren et al. 1997).  We have therefore not 
considered this case of mass transfer episodes in our models.

\end{enumerate}

\begin{figure}
  \plotfiddle{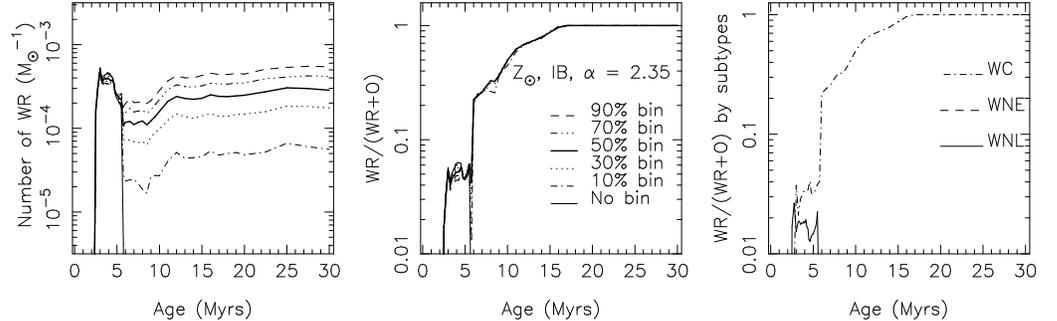}{4cm}{0.}{73.}{73.}{-230.}{-170.}
  \caption[]{(a) Predicted number of WR stars, normalized by the mass of
    stars formed in the cluster.  (b) WR over WR plus O stars;
    note that the ratio is almost independent on the binary abundance.  (c)
    WR types. }
  \label{WRf}
\end{figure}

\subsubsection{$\bullet$ Gainer star.}
We have assumed that the gainer star will be able to accrete the mass lost
by the donor if it is still in the Main Sequence, with a convective
envelope. In such cases we have considered that 50\% of the mass and
angular momentum of the system is lost anyway by the system in the mass
transfer process (Meurs \& van den Heuvel 1989). The gainer will become a
more massive star, which will evolve as a younger one. Combining the
evolutive status of the star before the RLOF episode and the amount of mass
accreted, we compute a time correction and make the star to evolve
following the track of a younger, more massive star. The net effect of
these processes will be the ``rejuvenation'' of the cluster (Van Bever \&
Vanbeveren 1998), which will host massive stars at ages where all of them
should have already exploded as SN, if only single stars where present.

If the gainer is in more evolved phases when RLOF occurs, we have assumed
that it will not be able to accrete mass, and that all the mass transferred
is lost by the system. In these cases common envelope evolution and
spiral-in processes might become important. 

\subsection{SN explosions and formation of High Mass X-Ray Binaries}

After a supernova explosion, the binary system will survive only if the
mass of the non-exploding star is two times larger than the mass of the
exploding one. We test this conditions when each of the massive stars in
the synthetic cluster exhaust its nuclear fuel, and compute then the number
of systems that will remain bounded or unbounded after the SN explosion.
If the exploding star had at the moment of the explosion a mass higher
than 8~M$_\odot$, we assumed that it becomes a black hole with more than 
4~M$_\odot$. If the mass is lower than 8, but higher than 4 M$_\odot$, it 
will become a neutron star with around 3~M$_\odot$. Otherwise, it will
become a White Dwarf with 1.4 M$_\odot$.
After the SN explosion the system might become very eccentric, depending on
its mass ratio and orbital separation before the explosion. We have assumed
that this eccentric post-SN systems might become a Be/X star (highly
eccentric systems with a Main Sequence star plus a compact companion).  As
the remnant star evolves, it will become a Supergiant. Then we assume that
the orbit will be circularized, giving rise to a ``permanent'' High Mass
X-Ray Binary (HMXRB).

\begin{figure}
  \plotfiddle{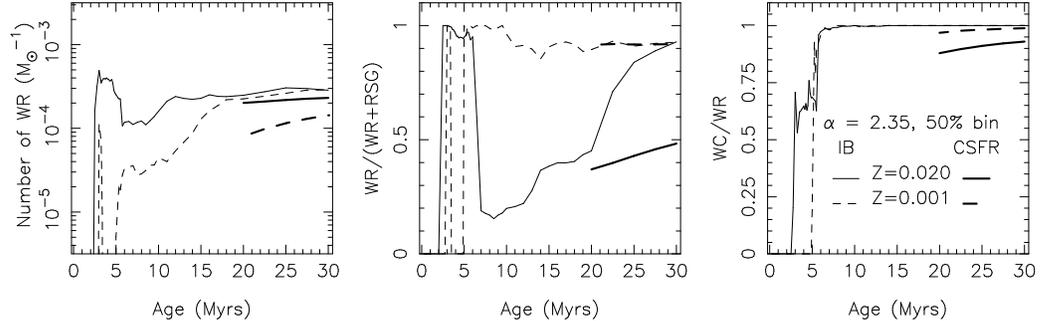}{4cm}{0.}{73.}{73.}{-230.}{-170.}
  \caption[]{(a) Predicted WR population for two metallicities.
    (b) WR over WR~+~RSG ratio. (c) WC over WR ratio. Predictions for
    continuous star formation rate (CSFR) have also been included.}
  \label{WRz}
\end{figure}

\section {Model predictions}

We will summarize shortly in this section the most relevant predictions of
our evolutionary models affected by the presence of binary systems. 

\subsubsection{$\bullet$ Wolf-Rayet stars population.}  
One of the most relevant effects of the mass transfers processes in close
binaries is the generation of WR stars at ages where models based on single
stars only do not predict the presence of any WR at all. We show in
Fig.~\ref{WRf} the predicted number of WRs, as well as their types and
ratios over O stars, as a function of the binary frequency. It can be seen
that while the generation of WR stars from the evolution of single stars
remains essentially the same, after around 5 Myr a relatively large number
of WR stars is originated in close binary systems. We want to stress that
massive stars in detached binary systems will evolve essentially as single
stars, and that this new population of WR stars is related only to the
binary systems experiencing interactions.

In Fig.~\ref{WRz} we show the dependence of the WR population properties on
metallicity. It is interesting to see that the formation of WRs by the
binary channel is also metallicity dependent, as a consequence of the
different evolution of single stars at lower metallicities. An important
result of our models is the prediction of the simultaneous presence of WR
and Red Supergiant (RSG) stars in evolved (older than 7~Myr) clusters
formed at metallicities above around Z=0.008. 

Finally, it is also important to see that our models predict that the WR
stars formed by the binary channel will be preferentially of the WC
type. As explained above, WR stars formed in Case~B will loose their
envelope and will show a surface abundance similar of that in the nucleus,
i.e., rich in Carbon.

\subsubsection{$\bullet$ OB stars and $W(H\beta)$.} The accretion of mass by
transfer episodes will originate continuously a population of relatively
massive stars at ages where they should have already disappeared from the
young clusters (assuming nearly instantaneous bursts). These stars,
together with the hot WR stars formed, will maintain the ionizing power of
the clusters at a significantly higher level than just single stars. As a
result, the strength of the emission lines will decrease with time much
slower. With a binary frequency of 50\%, $W(H\beta)$, frequently used as an
age estimator, will be larger by more than an order of magnitude at 15~Myr
than the value predicted for single stars clusters.

\subsubsection{$\bullet$ Systems with compact objects.} The fast evolution 
of the massive primary originates the formation of close systems with a
compact (black hole, neutron star) component after the first 4--5 Myr of
evolution. The accretion of mass onto the surface of the compact companion
will give rise to high-energy phenomena. As an example, we predict around
10 HMXRB to be present in a 15~Myr cluster having formed around
10$^7$~M$_\odot$ of stars. 

\subsubsection{$\bullet$ Spectral Energy Distribution (SED).} The SED will
be drastically affected by the presence of binary systems in the
cluster. The formation of HMXRB and related objects provides a source of
high-energy emission which completely dominates the hard X-ray range (above
around 2~keV) after the first 4--5 Myr. More details are given in the
accompanying contribution by Cervi\~no \& Mas-Hesse (this volume).

\acknowledgments

This work has been  supported by Spanish CICYT ESP-95-0389-C02-02 and INTA
``Rafael Calvo Rod\'es'' grants.


\begin{references}

\reference
Arnault, Ph. 1990, Ph.D. Thesis, IAP (France)

\reference
Cervi\~no, M. and Mas-Hesse, J.M. 1994, \aap, 284, 749 (CMH)

\reference
Cervi\~no, M., Mas-Hesse, J.M. and Kunth, D.
1996, in ``WR star in the Framework of Stellar Evolution'', 33rd Li\`ege 
International Astrophysical Colloquium, p. 613

\reference
De Greve, J.P. and de Loore, C.W.H.
1992, \aaps,  96, 653

\reference
Gonz\'alez-Delgado, R., Heckman, T., Leitherer, C., Meurer, G., Krolik, J.,
Wilson, A.S., Kinney, A. and Koratkar, A. 1998, \apj, 505, 174

\reference
Heckman, T.M., Gonz\'alez-Delgado, R., Leitherer, C., Meurer, G., Krolik, J., 
Kinney, A.,  Koratkar, A., Wilson, A.S.   1997, \apj, 482, 114


\reference
Langer, N.
1989, \aap, 210, 93

\reference Maeder, A. and Meynet, G.
1994, \aap, 287, 803

\reference
Mas-Hesse, J.M. and Kunth, D. 1991, \aaps, 88, 399 (MHK)

\reference
Meurs, E.J.A. and van den Heuvel, E.P.J.
1989, \aap, 226, 88

\reference
Schaerer, D. and Maeder, A.
1992, \aap, 263, 129

\reference Schaerer, D. and Vacca, W.D. 1998, \apj, 497, 618

\reference 
Schaller, G., Schaerer, D., Meynet, G. and Maeder, A.
1992, \aaps, 96, 269

\reference 
Van Bever, J. and Vanbeveren, D. 1998, \aap, 334, 21 

\reference 
Vanbeveren, D. and Conti, P.S.
1980, \aap,  80, 230

\reference
Vanbeveren, D. 1991, \aap, 252, 159

\reference 
Vanbeveren, D., Van Bever, J. and de Donder, E. 1997, \aap, 317, 487

\reference 
Vanbeveren, D., de Donder, E., Van Bever, J., Van Rensbergen, W. and de 
Loore, C.W.H.
1998, New Astronomy 3, 443

\end{references}
\end{document}